\documentclass[prl,aps,twocolumn,showpacs,floatfix,superscriptaddress]{revtex4}
\usepackage{graphicx}
\usepackage{bm}
\usepackage{amsmath}
\usepackage{amssymb}
\usepackage{amsfonts}
\usepackage{enumitem}

\begin{document}

\title{Silo collapse under granular discharge}
\author{G. Guti\'errez}
\affiliation{Departamento de F\'isica, Universidad Sim\'on Bol\'ivar, Apdo. 89000, Caracas 1080-A, Venezuela}
\affiliation{PMMH, ESPCI, CNRS (UMR 7636) and Univ. Paris 6 \& Paris 7, 75005 Paris, France}
\author{C. Colonnello}
\affiliation{Departamento de F\'isica, Universidad Sim\'on Bol\'ivar, Apdo. 89000, Caracas 1080-A, Venezuela}
\author{P. Boltenhagen}
\affiliation{UMR CNRS 6251, Universit\'e de Rennes 1, 35042 Rennes Cedex, France}
\author{J. R. Darias}
\affiliation{Departamento de F\'isica, Universidad Sim\'on Bol\'ivar, Apdo. 89000, Caracas 1080-A, Venezuela}
\author{R. Peralta-Fabi}
\affiliation{PMMH, ESPCI, CNRS (UMR 7636) and Univ. Paris 6 \& Paris 7, 75005 Paris, France}
\affiliation{Departamento de F\'isica, Facultad de Ciencias, Universidad Nacional Aut\'onoma de M\'exico, 04510, M\'exico D.F., M\'exico}
\author{F. Brau}
\affiliation{Nonlinear Physical Chemistry Unit, Universit\'e Libre de Bruxelles (ULB), CP231, 1050 Brussels, Belgium}
\author{E. Cl\'ement}
\affiliation{PMMH, ESPCI, CNRS (UMR 7636) and Univ. Paris 6 \& Paris 7, 75005 Paris, France}

\begin{abstract}
We investigate, at a laboratory scale, the collapse of cylindrical shells of radius $R$ and thickness $t$ induced by a granular discharge. We measure the critical filling height for which the structure fails upon discharge. We observe that the silos sustain filling heights significantly above an estimation obtained by coupling standard shell-buckling and granular stress distribution theories. Two effects contribute to stabilize the structure: (i) below the critical filling height, a dynamical stabilization due to granular wall friction prevents the localized shell-buckling modes to grow irreversibly; (ii) above the critical filling height, collapse occurs before the downward sliding motion of the whole granular column sets in, such that only a partial friction mobilization is at play. However, we notice also that the critical filling height is reduced as the grain size, $d$, increases. The importance of grain size contribution is controlled by the ratio $d/\sqrt{R t}$. We rationalize these antagonist effects with a novel fluid/structure theory both accounting for the actual status of granular friction at the wall and the inherent shell imperfections mediated by the grains. This theory yields new scaling predictions which are compared with the experimental results.
\end{abstract}

\pacs{83.80.Hj,47.57.Gc,47.57.Qk,82.70.Kj}

\date{\today}
\maketitle

Granular media are ubiquitous in food industry, agriculture, pharmacy, chemistry and construction, to name a few. This state of matter is the subject of intense research to understand its complex and diverse properties (flow, rheology, patterns, etc.)~\cite{jaeg96,PGDG98,camp06,aran09,andreotti13}. Cylindrical containers are frequently used to store granular material. Silo collapses resulting of faulty constructions or undetected structural deterioration are particularly vicious industrial accidents~\cite{doga09,dutt13}. Each year, in spite of severe regulations defining the design and the use of granular storage devices, dramatic financial and human tolls stem from unexpected structural breakdown. Failure in reducing significantly such risks point on fundamental difficulties in accounting properly for the thin shell structural properties and the physics of granular matter altogether. There is a vast engineering literature on buckling instabilities of empty shells~\cite{timo61,yama84,teng96,sing97,sing02,teng03} but thin cylindrical shells filled with grains constitute a more complex physical system that is largely unresolved~\cite{rott09} and its investigation leads to open problems of great practical and scientific interest~\cite{brow98,chen08}.

A common failure mode in cylindrical metal silos is the buckling under axial compression which is often triggered by gravity driven discharges of granular material~\cite{rott09}. Here we propose a systematic study of this problem based on laboratory scale silos. The  conditions under which such silos collapse during discharge are investigated as a function of various parameters characterizing both the silo and the grains. This study is supported by a theoretical approach that couples, in the simplest possible way, the theory of buckling of thin shells and the presence of a granular material on the inside, as a source of possible imperfections. 

\begin{figure}
\includegraphics[width=\columnwidth]{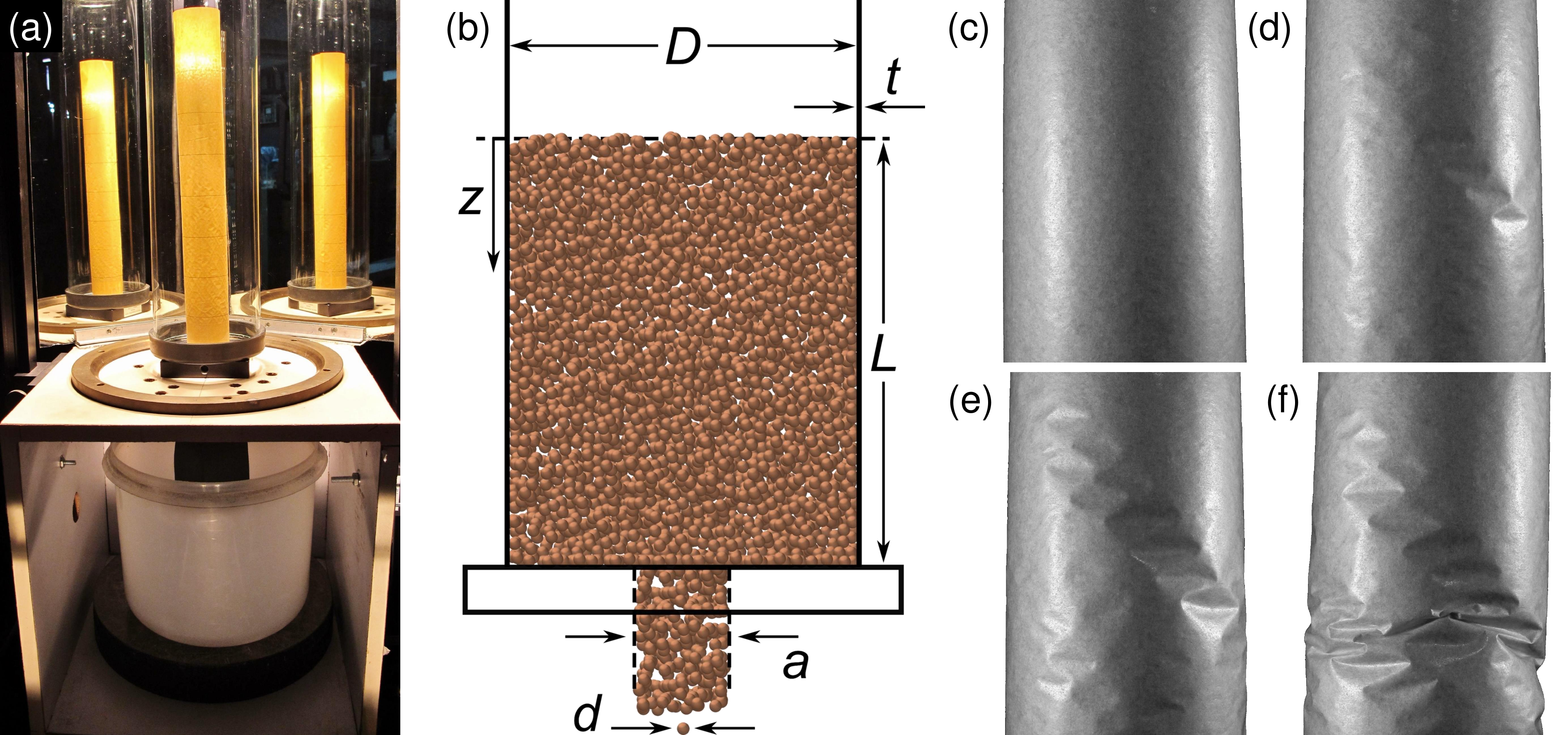}
\caption{(color online) Experimental set-up and buckling sequence: (a) Picture of the experimental apparatus showing a paper silo, the filling device and the two mirrors used for complete visualization of the discharge. (b) Schematic cross-section of the silo with its dimensions. (c-f) Time sequence of deformations in a collapsing silo during grain discharge. Glass beads $d =1.5\pm 0.1$ mm, column thickness $t = 27\pm 5 $ $\mu$m and $R=D/2=2.00\pm 0.05$ cm.}
\label{Fig1}
\end{figure}

The experimental set up is simple in its principle. A thin paper cylindrical shell is filled with granular material of size $d$ and density $\rho$ up to a certain height $L$. The silo is then emptied through a bottom circular aperture of diameter $a=2.50 \pm 0.05$ cm which is closed with a plug during filling. The conditions of discharge are recorded by two video cameras: One recording the motion of the grains in the upper part of the silo, and a second one providing a global vision of the silo. Furthermore, we placed behind the silo two mirrors making an angle of $45^{\circ}$ with respect to the viewing direction to provide a vision of the whole silo circumference. In these experiments, $a/d>5$ to ensure continuous granular flow during the discharge~\cite{zuri03,mank07}. Silos of different radii $R$, and thickness $t$, are prepared using a paper sheet wrapped around a metal tube and glued along a narrow band to form a cylindrical shell. The ratio $R/t$ investigated is compatible with some industrial steel silos, for which $300<R/t<3000$~\cite{brow98}; however we kept the grain size much larger than the wall thickness, $d\gg t$. Since paper is an anisotropic material, the silos were prepared using the same orientation of the sheet of paper and the cylindrical axis. The shell, which is inserted into a rigid cylindrical base, is fixed at the bottom and left free at the top. The preparation protocol is strict, to avoid any residual twist that would affect the shell mechanical strength. The paper Young modulus in the silo vertical direction has been measured by flexural tests ($E=2\pm 1$ GPa). Figure~\ref{Fig1}(a-b) shows a picture and a schematic diagram of the experimental setup used to determine the collapse height, $L_{\text{c}}$, under grain discharge.

Most experiments were performed using spherical beads with a diameter larger than 1 mm to reduce the relative importance of disturbances, such as humidity or electrostatic interactions, with respect to gravity forces. The height $L$ of the granular column is gradually increased after each successive full discharge until the silo collapses for $L=L_{\text{c}}$ during the final discharge. Figure~\ref{Fig1}(c-f) shows four successive snapshots of a silo after the discharge onset, for $L>L_{\text{c}}$, such that a collapse occurs. We observe how initial diamond shaped deformations localized near the silo bottom (Fig.~\ref{Fig1}(d)) assemble into a cluster propagating upwards on the cylindrical surface (Fig.~\ref{Fig1}(e)) until a large plastic deformation develops followed by a collapse of the silo (Fig.~\ref{Fig1}(f)) (see also movie in supplemental material). 

Figure~\ref{Fig2}(a) displays the position of the upper layer of grains, $z$, as a function of time and measured relatively to the collapse height, $L_{\text{c}}$, during two identical experiments where $L$ is either below or above $L_{\text{c}}$. For $L<L_{\text{c}}$, localized diamond dimples may appear at the discharge onset, as shown by the dashed arrow on Fig.~\ref{Fig2}(d), but they are progressively smoothed out during the discharge such that the empty silo recovers its initial state. Two examples of this ``dynamical stabilization" process are shown in the supplementary movie. For $L>L_{\text{c}}$, irreversible plastic deformations of the silo occur before the end of the discharge process and the onset of collapse never occurs after the downwards sliding of the whole grain column (see Ref.~\cite{colo14} for a systematic study of this effect).  

\begin{figure}
\centering
\includegraphics[width=\columnwidth]{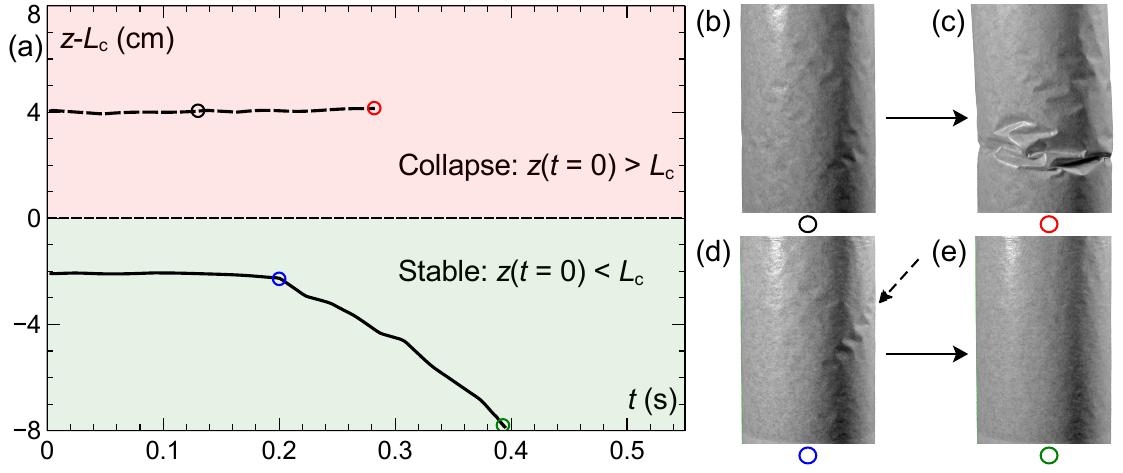}
\caption{(color online) Experiments where silos of radius $R=2.00 \pm 0.05$ cm and $t=27 \pm 5$ $\mu$m are filled by glass beads with $d=3.0 \pm 0.1 $ mm. (a) Position of the upper layer of grains $z$ as a function of time measured relatively to the collapse height, $L_{\text{c}}$ ($L=z(t=0)$). When $L>L_{\text{c}}$, irreversible deformations occur leading to a collapse of the silo. The circles indicate the time at which pictures (b-e) are taken. (b-c) Pictures of two states occurring during discharge onset for $L>L_{\text{c}}$. Panel~(c) shows a collapsed silo. (d-e) Pictures of two states occurring at the discharge onset and once the discharge is completed for $L<L_{\text{c}}$. The solid arrows indicate the temporal evolution and the dashed arrow shows localized diamond dimples smoothed out during the discharge.}
\label{Fig2}
\end{figure}

\begin{figure*}
\centering
\includegraphics[width=\textwidth]{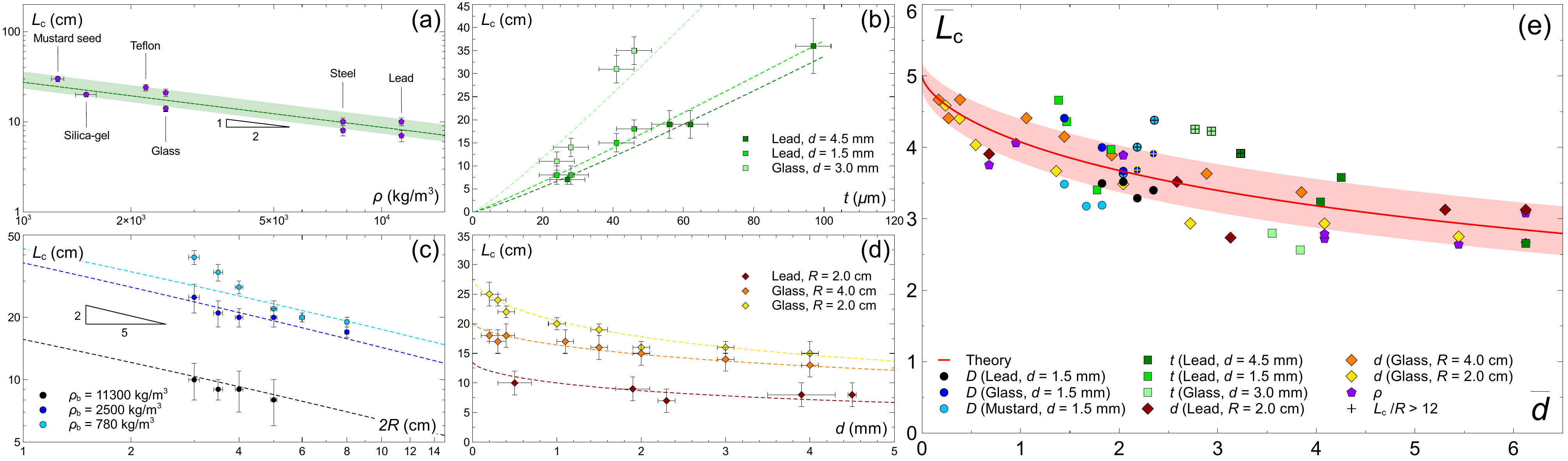}
\caption{(color online) Parametric exploration of the collapse height $L_{\text{c}}$ and theoretical outcome. (a) $L_{\text{c}}$ as a function of the grain density $\rho$ with $t=27\pm 5$ $\mu$m, $R=2.00\pm 0.05$ cm and $d \in [0.5,4.5]$ mm. The dashed line is obtained with Eq.~(\ref{critical-height}) and the shaded area represents the region spanned by varying $d$ in the experimental range. (b) $L_{\text{c}}$ as a function of the silo thickness $t$ (Lead, $d=4.5\pm 0.1$ mm, $R=2.00\pm 0.05$ cm / Lead, $d=1.5 \pm 0.1$ mm, $R=2.55\pm 0.05$ cm / Glass, $d=3.0 \pm 0.1$ mm, $R=2.55\pm 0.05$ cm). (c) $L_{\text{c}}$ as a function of the silo radius $R$ with $t=27\pm 5$ $\mu$m and $d=1.5 \pm 0.1$ mm (Lead / Glass / Mustard seed). (d) $L_{\text{c}}$ as a function of the grain size $d$ with $t=27\pm 5$ $\mu$m (Glass, $R=2.00\pm 0.05$ cm / Glass, $R=4.00\pm 0.05$ cm / Lead, $R=2.00\pm 0.05$ cm). (b-d) The dashed lines are obtained with Eq.~(\ref{critical-height}). (e) Rescaled collapse height $\bar{L}_{\text{c}}$ defined in Eq.~(\ref{critical-height}) as a function of the rescaled grain size $\bar{d}$ defined in Eq.~(\ref{Delta-final}). The solid line is obtained with Eq.~(\ref{critical-height}) and the shaded area corresponds to the region spanned when $\chi$ and $\gamma$ are varied ($\chi = 5.0 \pm 0.2$ and $\gamma = 0.11 \pm 0.03$).}
\label{Fig3}
\end{figure*}

A perfect elastic thin cylindrical shell of radius $R$ and thickness $t$ buckles under uniform axial compression when the applied stress exceeds the critical limit~\cite{timo61,yama84}
\begin{equation}
\label{critical-stress}
\sigma_{\text{c}}= \frac{Et}{\sqrt{3(1-\nu^2)}R},
\end{equation}
where $E$ and $\nu$ are the Young modulus and the Poisson ratio of the cylinder's isotropic elastic material. Axisymmetric or asymmetric buckling modes occur at the same critical stress. In our case, the applied load is not uniform and is due to granular material which exerts a shear force on the inner wall of the shell. This shear force, $F_{\mu}$, pushing down the structure is given by $F_{\mu}(z)= 2\pi R\int_{0}^{z}{\sigma_{rz}(z')\,dz'}$~\cite{supp}. Experiments have shown that the shear stress distribution $\sigma_{rz}(z)$ at the wall of a cylindrical column, is given by the so-called Janssen's stress profile with a good accuracy either in the static~\cite{ovar03} or in the dynamic case~\cite{bert03}:
\begin{equation}
\label{shear-stress}
\sigma_{rz}(z)=K\mu_{\text{w}} \,\rho_{\text{g}} g \,\lambda\left(1-e^{-\frac{z}{\lambda}}\right),
\end{equation}
where $\mu_{\text{w}}$ is the grain-wall Coulomb static friction coefficient, $K$ is an effective vertical to horizontal redirection coefficient and $\lambda=R/2K\mu_{\text{w}}$ is the Janssen's screening length. $\rho_{\text{g}}$ is the bulk density of the granular medium and is related to the density of the grain material, $\rho$, through the packing fraction $\varphi$ ($\rho_{\text{g}}=\varphi \rho$); $\varphi\simeq 0.64$ for random close packed spheres~\cite{finn70}. When all the contact shear forces at the wall are polarized upwards, the screening length $\lambda$ is of the order of the column diameter $2R$~\cite{ovar03}. However, in general, just after pouring the grains, the mobilization status of the contact friction forces at the wall may depend on complicated dynamical processes, involving the pouring history~\cite{ovar03}. The saturation of the stress profile thus takes place over a larger distance from the top surface. Some models tentatively assume a random mobilization of the friction forces at the wall to describe this effect~\cite{ditl99}. In a simplified Janssen's picture this would be equivalent to a large value of the screening length $\lambda$, meaning that the wall bears less load than expected in the case of a full friction mobilization. However, when the discharge begins, the upward friction mobilization increases until the grains may move downwards. Janssen's profile can be recovered with a great accuracy provided a large amount of granular material is released during the discharge~\cite{ovar03, perg12}. In our experiments, we do not expect a full friction mobilization after pouring and collapse occurs at the discharge onset. Consequently, to model, in the simplest way, the fact that the upward friction polarization at the wall may not be achieved when the collapse occurs, we introduce an empirical dimensionless parameter $\xi$ such that $\lambda= \xi R$ where $\xi$ can be varied from $\xi = O(1)$ (full mobilization) to $\xi \gg L/R$ (random mobilization). 

The stability against axisymmetric buckling of a perfect elastic thin cylindrical shell subject to the shear force $F_{\mu}$ induced by granular material has been studied in detail in~\cite{supp}. However, in the limit $t\ll R$, the relevant scaling can simply be obtained by balancing the critical stress (\ref{critical-stress}) and the applied load ($F_{\mu}(L)=2\pi R\, t\, \sigma_{\text{c}}$). In the case of full friction mobilization, the shear force $F_{\mu}$ evolves essentially linearly with $z$ (except for $z\ll R$) whereas for random friction mobilization, it behaves as a quadratic function of $z$. The random mobilization regime can thus be view as a pseudo hydrostatic regime, {\it i.e.} an hydrostatic regime with $\mu_{\text{w}}>0$. Both regimes yield quite distinct scaling:
\begin{subequations}
\begin{align}
\label{full-mob}
F_\mu(z)  \simeq \pi R^2 \rho_{\text{g}} g\, z, \quad \xi \lesssim 1 \quad &\Rightarrow   L_{\text{c}} \simeq \frac{Et^2}{\rho_{\text{g}}g R^2}, \\
\label{random-mob}
F_\mu(z)  \simeq \frac{\pi R \rho_{\text{g}} g}{2\xi}\, z^2, \quad \xi \gg \frac{L}{R} \quad &\Rightarrow   L_{\text{c}} \simeq \sqrt{\frac{Et^2}{\rho_{\text{g}}g R}}.
\end{align}
\end{subequations}
However, even in regimes where the Taylor expansion (\ref{random-mob}) of $F_{\mu}$ is not justified, the shear force can still be approximated to a good accuracy by a quadratic function for $z\in [0, \sim 1.6\, \xi]$~\cite{supp}. Consequently, while a pseudo hydrostatic regime takes place strictly only for $L/R \ll \xi$, an effective pseudo hydrostatic regime applies for $L/R$ as large as $\sim 1.6\, \xi$ and leads in good approximation to the scaling (\ref{random-mob}).

Figure~\ref{Fig3}(a-d) shows a parametric study of the collapse heights $L_{\text{c}}$ for various grain parameters ($\rho$ and $d$) and silo parameters ($R$ and $t$). Figure~\ref{Fig3}(a-b) clearly indicate that $L_{\text{c}}\sim t/\rho^{1/2}$ which is only compatible with the scaling (\ref{random-mob}). The scaling with $R$ reported in Fig.~\ref{Fig3}(c) is also incompatible with (\ref{full-mob}) and it is consistent with Eq.~(\ref{random-mob}) (and improved by Eq.~(\ref{critical-height})); this supports the contention that full mobilization is not occurring. We recall that the sliding of the granular material produces an increase of the upward friction mobilization and a classical Janssen's stress distribution at the wall should be obtained. The parameter $\xi$, describing the friction mobilization, should thus vary from large to $O(1)$ values. The various scaling reported in Fig.~\ref{Fig3}(a-c) indicate that, even if $\xi$ may decrease during this process, it never reaches values of order~1. These results confirm recent measurements proving that the discharge of a large amount of granular material is necessary to reach $\xi \sim 1$~\cite{perg12} whereas the collapse of our silos occurs at the discharge onset (see Fig.~\ref{Fig2}).

Figure~\ref{Fig3}(d) shows that $L_{\text{c}}$ depends also on the grain size $d$ which is not a parameter of the model so far because the granular mater is assumed to be continuous in the derivation of the shear stress, Eq.~(\ref{shear-stress}), in agreement with experiments for $d/2R\lesssim 0.1$~\cite{camb13,qadi10}. Therefore, neither the grain size nor the rigidity of the cylinder affect the stress distribution in the regimes considered here~\cite{camb13,qadi10}. One must then search for another explanation for the dependence of $L_{\text{c}}$ on $d$. In the regime $d\gg t$ we consider, the silo wall is slightly deformed once it is filled with the grains. The shell is thus not a perfect cylinder at the discharge onset. We propose to describe these imperfections as geometric imperfections even if they are not necessarily stress free. An asymptotic formula, valid for small magnitude of the imperfections, has been developed in Ref.~\cite{amaz72} to describe axisymmetric geometric imperfections and was tested in Ref.~\cite{hutc71}. If an initial localized axisymmetric imperfection $\omega_0$ is present on the surface of a cylindrical shell of constant thickness $t$, the stress, $\sigma$, for which the shell buckles under axial compression is a solution of 
\begin{equation}
\label{P-imperfect}
(1-\sigma/\sigma_{\text{c}})^{3/2} = \eta (\sigma/\sigma_{\text{c}}),
\end{equation}
where $\sigma_{\text{c}}$ is the classical buckling stress, Eq.~(\ref{critical-stress}). The expression for $\eta$ is determined by the shape of the imperfection as follow
\begin{equation}
\label{Delta}
\eta =\frac{3^{\frac{3}{2}}|\Delta|}{2} , \quad \Delta= \sqrt{\frac{1-\nu^2}{2}} \int_{-\infty}^{\infty} \frac{w_0(\tilde{x})}{h} e^{i \tilde{x}} d\tilde{x},
\end{equation}
with $\tilde{x} = \pi x/\lambda_{\text{c}}$ and $\lambda_{\text{c}} = \pi \sqrt{Rt}/[12(1-\nu^2)]^{1/4}$ (the half wavelength of the classical axisymmetric buckling mode). Equation (\ref{P-imperfect}) admits a simple accurate approximate solution for small $\Delta$ where this asymptotic formula applies~\cite{supp}:
\begin{equation}
\label{sol-delta}
\sigma\simeq \sigma_{\text{c}} (1+|\Delta|^{2/3})^{-2}.
\end{equation}
Let us consider a localized imperfection $\omega_0 = \delta f(x)$ with $f(0)=1$, $|f(x)| \le 1$ and where $f(x)$ is vanishing for $|x| > \lambda_{\text{d}}$. At the first order in $\pi \lambda_{\text{d}} / \lambda_{\text{c}}$ we obtain~\cite{supp}: $\Delta = \pi \sqrt{2(1-\nu^2)}\, (\delta \lambda_{\text{d}}/ t \lambda_{\text{c}})$. Assuming that the amplitude of the imperfection is of order of the thickness $t$ and that its spatial extension is of order of the grain size $d$, we obtain
\begin{equation}
\label{Delta-final}
\Delta = \gamma \frac{d}{\sqrt{R t}} = \gamma \bar{d},
\end{equation}
where $\gamma$ is a constant fixed by the experimental data. Balancing the modified critical stress Eq.~(\ref{sol-delta}) with the applied load $F_\mu(L_{\text{c}})$ given by Eq.~(\ref{random-mob}) and using Eq.~(\ref{Delta-final}) we get the collapse height
\begin{equation}
\label{critical-height}
\bar{L}_{\text{c}}=\frac{\chi}{1+ (\gamma \bar{d})^{\frac{2}{3}}}, \quad \text{with} \quad \bar{L}_{\text{c}} = \frac{L_{\text{c}}}{t} \sqrt{\frac{\rho g R}{E}},
\end{equation}
and where $\chi=\sqrt{2\xi/\varphi}\, (1-\nu^2)^{-1/4}$~\cite{supp}. Comparison of this relation (dashed lines) with the data presented in Fig.~\ref{Fig3}(a-d) shows a good agreement provided $\chi = 5.0 \pm 0.2$ and $\gamma = 0.11 \pm 0.03$. 
Notice that the grain-wall friction coefficient, $\mu_{\text{w}}$, the grain-grain friction coefficient, $\mu_{\text{g}}$, and $\xi$ are related by $\mu_{\text{g}}=(2\mu_{\text{w}}\xi-1)/2\sqrt{2\mu_{\text{w}}\xi}$~\cite{supp}. From the range of possible values for $\xi$, we find $\mu_{\text{g}}\simeq 0.60 \pm 0.06$ for $\mu_{\text{w}}\simeq 0.2$ which is quite reasonable.

The data reported in Fig.~\ref{Fig3}(b) indicate, by simple linear extrapolation, that the collapse height vanishes for an ``apparent" finite thickness $t_{\text{c}}\simeq 10$ $\mu$m. This value is about two order of magnitude larger than the thickness at which the silos would collapse under their own weight. Actually, the expression (\ref{critical-height}) of $L_{\text{c}}$ is a convex function of $t$. This apparent critical thickness, $t_{\text{c}}$, can be obtained from the model by extrapolating the asymptote of Eq.~(\ref{critical-height}) down to a vanishing $L_{\text{c}}$. Focusing on the dependence on $t$, Eq.~(\ref{critical-height}) can be written as follows together with its asymptotic expansion 
\begin{equation}
\label{critical-height-asymp}
L_{\text{c}}= \frac{p t}{\left[1+ \left(\frac{q}{t}\right)^{\frac{1}{3}}\right]} \underset{t\gg q}{\simeq} p \left(t^{\frac{2}{3}}+q^{\frac{2}{3}}\right)\left(t^{\frac{1}{3}}-q^{\frac{1}{3}}\right),
\end{equation}
where $q= (\gamma d)^2/R$. The asymptotic expansion vanishes for $t = q$ which corresponds to the apparent critical thickness: $t_{\text{c}} =(\gamma d)^2/R$. With the parameters used in the experiments reported in Fig.~\ref{Fig3}(b), one finds $1$ $\mu$m $<t_{\text{c}} < 12$ $\mu$m in good agreement with the value found by extrapolating the experimental data.

Finally, the data have been rescaled according to the scaling obtained in the model and gathered in Fig.~\ref{Fig3}(e). We observed a nice collapse of the data and a good agreement with Eq.~(\ref{critical-height}). As mentioned above, a pseudo hydrostatic regime is expected to apply for $L_{\text{c}}/R \lesssim 1.6\, \xi \simeq 12$. Data which do not satisfy this inequality are marked by a cross in Fig.~\ref{Fig3}(e).

In summary, we present an experimental study of granular discharge out of a thin cylindrical shell revealing paradoxical effects. The specific nature of granular matter displays contradictory stabilizing or destabilizing features controlling the structural collapse. On one hand, granular wall friction is stabilizing the structure and the columns can be filled up to levels significantly higher than what would be expected from elementary mechanical arguments. The reason is two-fold: (i) pouring processes reduce the shear stress at the walls due to a lack of friction mobilization, (ii) after the discharge onset, when wall friction is fully mobilized, the localized buckling structures do not grow up to collapse because of a dynamical stabilization induced by the granular flow. Therefore, the collapse is due to the possibilities for buckling patterns to grow unbounded, in the static phase prior to a full friction mobilization at the wall. On the other hand, the finite size nature of the grains create inherent noisy patterns impinging the shell stability. To account for the filling height above which collapse is observed, we presented a theoretical analysis accounting for the lack of initial friction mobilization at the wall and granular born defects in the shell. This leads to an original scaling behavior involving all the essential mechanical and geometrical parameters and to a good quantitative agreement with the experimental outcomes. 
The complex ``fluid"-structure problem studied here was addressed by explicit analytical methods using a minimal model which captures the essential physics associated with granular matter. The formalism could be extended to other wall geometries or shell structures. We show that such a fluid/structure problem is not only driven by forces applied during the flow but that it can also be influenced by unavoidable deformations occurring before any flow takes place. More generally, our results can have implications to other fluid/structure problems involving for example, advanced drilling techniques~\cite{Sadeghi2014}, plant root growth in sandy soils~\cite{Clark2003}, mobility of living organisms in sand~\cite{Goldman2014} and finally, the emergent field of ``soft-robotics"~\cite{Brown2014} which uses the interplay between granular matter and elastic membranes to perform various tasks.
 
\vspace{0.2cm}

\begin{acknowledgments}
We thank L.I. Reyes, I.J. S\'anchez, B. Roman and J.E. Wesfreid for useful discussions. We thank the PCP cooperation program, CNRS and Fonacit for their support. R.P.-F. enjoyed a Sabbatical Fellowship from the Direccion General de Asuntos del Personal Academico-UNAM, and the hospitality at the PMMH (ESPCI). F.B. thanks PRODEX for financial support. This work is funded by the ANR JamVibe and a CNES grant.
\end{acknowledgments}

\clearpage

\begin{center}
{\bf {\it Supplemental Material for}\\ ``Silo collapse under granular discharge"}
\end{center}

\section{Granular load and self-weight}

During the experiments, two forces are acting on the cylinder wall. One results from the interaction between the granular material and the wall through friction and the other is the cylinder self-weight. 

\subsection{Granular load}
\label{sec:gran-load}

\begin{figure}
\includegraphics[width=\columnwidth]{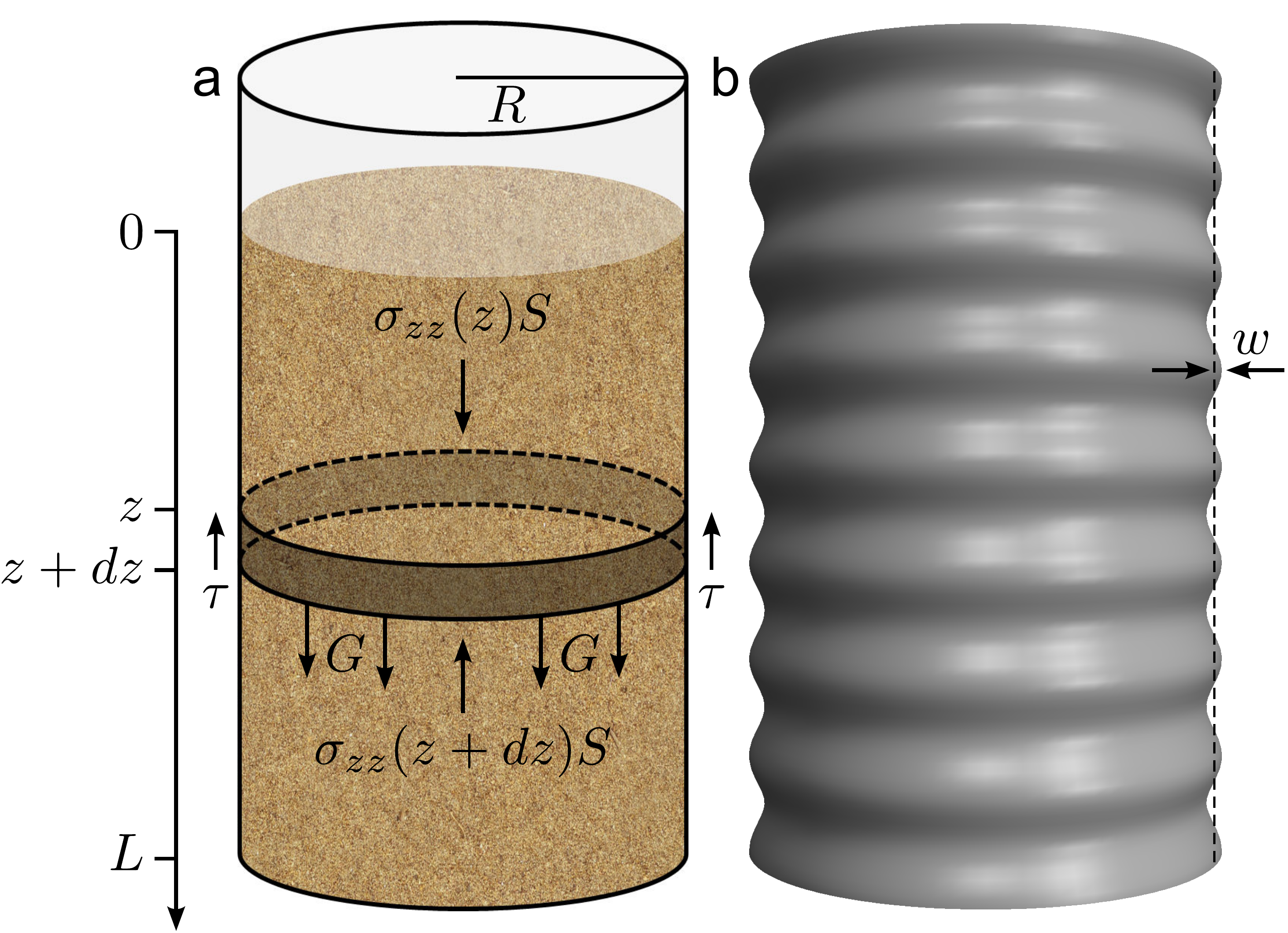}
\caption{{\bf a.} Schematic view of the forces acting on a thin layer of infinitesimal thickness $dz$ of granular material. {\bf b.} Schematic view of the axisymmetric buckling mode with the horizontal displacement $w$.}
\label{fig01}
\end{figure}

To compute the force due to the granular medium, we use the standard Janssen model~[1]. The assumptions underlying this model are:
\begin{itemize}
\item The vertical stress $\sigma_{zz}(z)$ is uniform along the cylinder section such as the resulting force is simply $\pi R^2 \sigma_{zz}(z)$ ;
\item Due to the friction $\mu_{\text{w}}$ between the grains and the wall, there exists a vertical upward force tangential to the wall, $\tau = \mu_{\text{w}} \sigma_{rr}(z) 2\pi R dz$, where $\sigma_{rr}(z)$ is the horizontal normal stress at the wall;
\item The horizontal normal stress is proportional to the normal vertical stress: $\sigma_{rr} = K \sigma_{zz}$, where $K$ is a constant (for a fluid $K=1$).
\end{itemize}
With those hypothesis, we can write the equilibrium equation for a thin layer of infinitesimal thickness $dz$ of granular material as depicted in Fig.~\ref{fig01}a:
\begin{equation}
\pi R^2\left(\sigma_{zz}(z)-\sigma_{zz}(z+dz)\right)-\tau + G = 0,
\end{equation}
with
\begin{eqnarray}
\label{tau1}
\tau &=& \mu_{\text{w}}\, \sigma_{rr}\, 2\pi R dz = \mu_{\text{w}} K\, \sigma_{zz}\, 2\pi R dz, \\
\label{g}
G &=& \rho_{\text{g}} g \pi R^2 dz,
\end{eqnarray}
where $G$ is the weight of the layer of granular material. $\rho_{\text{g}}=\varphi \, \rho$ is the density of the granular medium where $\rho$ is the grain density and $\varphi\simeq 0.64$ is the packing fraction for random close packed spheres~[2-5]. In the limit of vanishing $dz$, we obtain the following differential equation
\begin{equation}
\frac{d\sigma_{zz}}{dz}= \rho_{\text{g}} g - \frac{2\mu_{\text{w}} K}{R} \sigma_{zz},
\end{equation}
which admits the solution (knowing that $\sigma_{zz}(0)=0$),
\begin{equation}
\label{sigmazz}
\sigma_{zz}(z)=\rho_{\text{g}} g \lambda \left(1-e^{-z/\lambda}\right), \quad \text{with}\quad \lambda = \frac{R}{2\mu_{\text{w}} K} = \xi R.
\end{equation}

By definition, the force acting on the cylinder wall in contact with the thin granular layer of thickness $dz$ is 
\begin{equation}
\label{tau2}
\tau=\sigma_{rz} 2\pi R dz.
\end{equation}
The force acting on the entire wall, $F_\mu$, is thus the sum of these elementary forces
\begin{equation}
F_\mu = 2\pi R \int_0^z \sigma_{rz}(z') dz'.
\end{equation}
From the comparison between Eq.~(\ref{tau1}) and Eq.~(\ref{tau2}), we obtain
\begin{equation}
\sigma_{rz}(z)  = \mu_{\text{w}} K \sigma_{zz}(z),
\end{equation} 
which allow to compute $F_\mu$ explicitly with the help of Eq.~(\ref{sigmazz})
\begin{eqnarray}
\label{f-def}
F_\mu &=& 2\mu_{\text{w}} K \pi R \int_0^z \sigma_{zz}(z')  dz', \nonumber \\ 
&=& \pi R^3 \rho_{\text{g}} g \left(\bar{z}-\xi+\xi  e^{-\bar{z}/\xi} \right)= \pi R^3 \rho_{\text{g}} g\, {\cal F}(\bar{z},\xi) ,
\end{eqnarray}
where $\bar{z} = z/R\in [0,L/R]$.

Depending on the effective value of the parameter $\xi$ with respect to $L/R$ during the experiments, two asymptotic regimes are possible for the force $F_\mu$
\begin{eqnarray}
\label{janssen-limit}
F_\mu &\simeq& \pi R^2 \rho_{\text{g}} g\, z \quad \xi \ll 1 \quad \text{Janssen limit} \\
\label{hydro-limit}
&\simeq& \frac{\pi R \rho_{\text{g}} g}{2\xi}\, z^2 \quad \xi \gg L/R \quad \text{hydrostatic limit}
\end{eqnarray}
The relevance of these two asymptotic regimes for our experiments is discussed further in Secs.~\ref{sec:wkb} and \ref{sec:disc-hydro}.

\subsection{Self-weight}
\label{sec:self-w}

In our experiments, the force, $F_{\text{w}}$, induced by the weight of the silo wall is negligible compared to the granular load. Indeed, this force reads
\begin{equation}
\label{self-weight}
F_{\text{w}}=2\pi R t \rho_{\text{w}} g z,
\end{equation}
where $\rho_{\text{w}}$ is the density of the wall material. Comparison between Eqs.~(\ref{janssen-limit}), (\ref{hydro-limit}) and Eq.~(\ref{self-weight}) show that $F_{\text{w}} \ll F_{\mu}$ provided $t\ll 1-10$ cm (using typical values reported in Table~\ref{table1}). So, as anticipated, $F_{\text{w}}$ is negligible since in our experiments, $t$ does not exceed $100 \mu$m.

\section{Buckling of cylindrical shell}
\label{sec:1}

\subsection{Perfect shell}

\subsubsection{Constant load}

A perfect cylindrical shell buckles under axial compression when a constant compression stress applied on the cylinder, $\sigma$, exceeds a critical value, $\sigma_{\text{c}}$, given by~[6-9]
\begin{equation}
\label{classical-stress}
\sigma_{\text{c}} = \frac{E t}{\sqrt{3\left(1-\nu^2\right)}R},
\end{equation}
where $R$ and $t$ are the radius and thickness of the cylindrical shell and $E$ and $\nu$ are the Young modulus and the Poisson ratio the cylinder material respectively. This critical stress is the same for axisymmetric or asymmetric buckling modes.

\subsubsection{Granular load}

In our experiments, the granular force applies along the wall and depends on the $z$ coordinate. To treat this problem, we consider the axisymmetric buckling mode. Axisymmetric and asymmetric buckling mode share the same scaling with respect to the control parameter of the system. Moreover, if the asymmetric mode is of lower energy, the results obtained here can thus be considered as an upper bound on the critical buckling force. Neglecting the weight of the silo wall, the equation governing the silo stability against axisymmetric buckling is given by~[10]
\begin{equation}
\label{stability01}
B w'''' +\left(\frac{F_{\mu}}{2\pi R} w'\right)' +\frac{E t}{R^2} w=0, \quad B=\frac{Et^3}{12(1-\nu^2)},
\end{equation}
where $w$ is the horizontal displacement of the wall, see Fig.~\ref{fig01}b, $B$ is the bending modulus and $w'\equiv dw/dz$ with $z \in [0,L]$. If $F_{\mu}$ is below some critical value, the only solution satisfying homogeneous boundary conditions is $w=0$ and the silo is stable. The smallest value of $F_{\mu}$ for which a non vanishing solution exists is the critical buckling force. 

\begin{figure}
\includegraphics[width=\columnwidth]{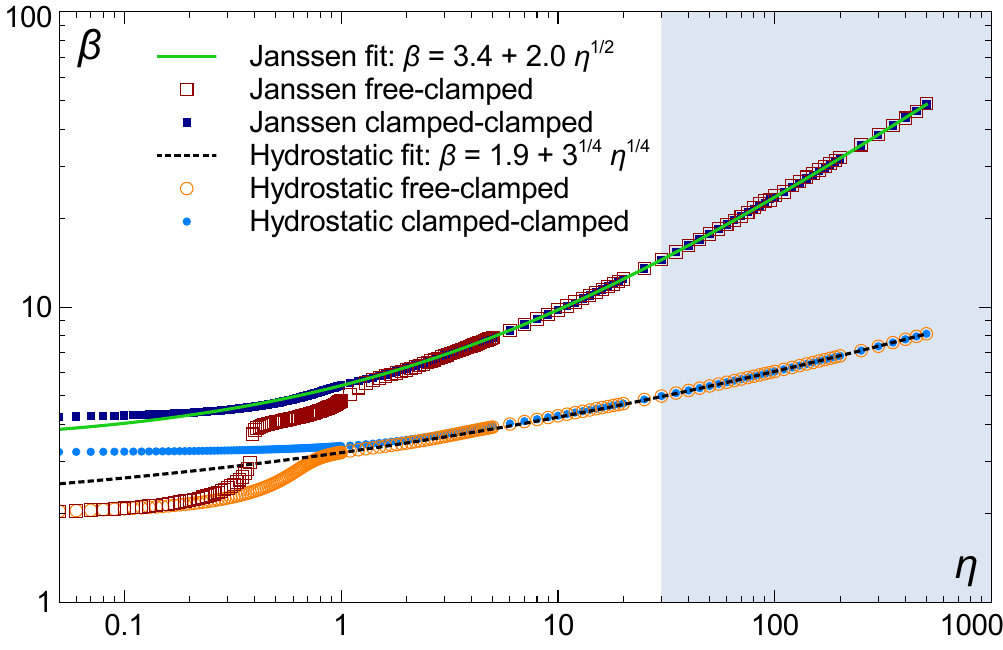}
\caption{Numerical solutions of Eq.~(\ref{stability02}) for the Janssen (\ref{janssen-limit}) and the hydrostatic (\ref{hydro-limit}) limits for the two boundary conditions (\ref{BC-FC}) and (\ref{BC-CC}). Solid and dotted lines show fits where the exponents are obtained from a WKB analysis. The relevant region for our experiments, $\eta >30$, is also indicated.}
\label{fig01b}
\end{figure}

We consider both limits (\ref{janssen-limit}) and (\ref{hydro-limit}) at once by defining
\begin{equation}
\label{c-def}
\frac{F_{\mu}}{2\pi R}= C_{\alpha} z^{\alpha}, \quad C_1 = \frac{1}{2}\rho_{\text{g}} g R, \quad C_2 = \frac{\rho_{\text{g}} g}{4\xi}.
\end{equation}
\begin{table}
\begin{ruledtabular}
\begin{tabular}{cccc}
\multicolumn{4}{c}{Silo}  \\ 
\hline
$\rho_{\text{w}}$ [kg/m$^3$] & $E$ [GPa s] & $t$ [$\mu$m] & $R$ [cm] \\ 
$\sim 667$ & $2\pm 1$ & $30-100$ & $1.5-4$\\
\hline \hline
\multicolumn{4}{c}{Granular material}  \\
\hline
$\rho$ [kg/m$^3$] & $\rho_{\text{g}}$ [kg/m$^3$] & $d$ [mm] & $L_{\text{c}}$ [cm]   \\
$1250-11340$ & $800-7250$ & $0.2-4.5$ & $7-39$  \\
\end{tabular}
\end{ruledtabular}
\caption{Typical values of the experiment parameters. $\rho_{\text{w}}$ and $E$ are the density and the Young modulus of the silo material respectively ($E$ is measured using a flexural test). $t$ is the wall thickness and $R$ the silo radius. $\rho$ is the grain density. $\rho_{\text{g}} = \varphi \rho$ is the granular medium density for a packing fraction $\varphi=0.64$. $d$ is the diameter of the beads and $L_{\text{c}}$ is the critical buckling height of the granular bed.}
\label{table1}
\end{table}
Applying the following change of variables
\begin{equation}
\label{beta}
z= \frac{L}{\beta}x, \quad \beta= L \left(\frac{C_{\alpha}}{B}\right)^{\frac{1}{2+\alpha}}, \quad x \in [0,\beta],
\end{equation}
Eq.~(\ref{stability01}) becomes
\begin{equation}
\label{stability02}
\ddddot{w}+x^{\alpha} \ddot{w}+\alpha x^{\alpha-1} \dot{w} + \eta w =0,
\end{equation}
with $\dot{w}=dw/dx$ and
\begin{equation}
\label{eta}
\eta = \frac{Et}{R^2}\left(\frac{B^{2-\alpha}}{C_{\alpha}^4}\right)^{\frac{1}{2+\alpha}}.
\end{equation}
We consider boundary conditions such as the bottom of the cylinder ($z=L \Rightarrow x=\beta$) is always clamped and the top of the cylinder ($z=x=0$) is either free or clamped:
\begin{align}
\label{BC-FC}
\text{Free--Clamped:}& \quad \left\{ \begin{matrix}
 \ddot{w}(0)=\dddot{w}(0)=0 &  \\
  w(\beta)=\dot{w}(\beta)=0 & 
 \end{matrix} \right. \\
\label{BC-CC}
\text{Clamped--Clamped:}& \quad \left\{ \begin{matrix}
 w(0)=\dot{w}(0)=0 &  \\
  w(\beta)=\dot{w}(\beta)=0 & 
 \end{matrix} \right.
\end{align}
Let's consider the boundary conditions (\ref{BC-FC}) to explain how the critical buckling force are obtained from Eq.~(\ref{stability02}). For a given value of $\alpha$, $\eta$ and $\beta$, Eq.~(\ref{stability02}) is solved by imposing three of the four homogeneous boundary conditions, for example $\ddot{w}(0)=\dddot{w}(0)=w(\beta)=0$, together with a fourth condition fixing the arbitrary amplitude (since Eq.~(\ref{stability02}) is linear) such as $w(0)=1$. With these four boundary conditions, there always exist a non vanishing solution of Eq.~(\ref{stability02}) but the fourth homogeneous condition ($\dot{w}(\beta)=0$) is satisfy only for specific value of $\beta$. Starting with a small value, the parameter $\beta$ is then increased until the fourth homogeneous boundary condition is satisfied ($\dot{w}(\beta)=0$). This particular value of $\beta$ together with Eq.~(\ref{beta}) gives the critical height of granular material above which the system is unstable. By varying $\eta$ for a given $\alpha$, one obtains the evolution of $\beta$ as a function of $\eta$. The results of this procedure are gathered in Fig.~\ref{fig01b}. We notice that above $\eta \gtrsim 1$, the influence of the specific boundary conditions used is negligible.

\subsubsection{WKB analysis and scaling for $\eta \gg 1$}
\label{sec:wkb}

To obtain additional exact results about the system, we perform the change of variable $z=L y$ ($y\in [0,1]$) to recast Eq.~(\ref{stability01}) under the following form
\begin{equation}
\label{stability03}
\varepsilon \ddddot{w}+ \psi \left(x^{\alpha} \ddot{w}+\alpha x^{\alpha-1} \dot{w} \right)+ w =0,
\end{equation}
where
\begin{equation}
\label{eps-phi}
\varepsilon = \frac{B R^2}{Et L^4} = \frac{1}{\eta \beta^4}, \quad \psi = \frac{L^{\alpha-2} C_{\alpha} R^2}{Et} = \frac{\beta^{\alpha-2}}{\eta}.
\end{equation}
In the limit $\eta \gg 1$, $\varepsilon$ becomes arbitrarily small and a standard WKB analysis can thus be applied~[11]. This allow us to obtain the exact scaling between $\beta$ and $\eta$ in the limit $\eta \gg 1$ relevant in our system. Indeed, from the typical values of the parameters used in our experiments and reported in Table~\ref{table1}, we note that $\eta \in [34,84500]$ for $\alpha=1$ and $\eta \in [2000,255000]$ for $\alpha=2$ which justifies the WKB approximation we used in this section.

For this purpose, we consider the following formal expansion
\begin{equation}
w(x) = e^{\frac{1}{\delta} \sum_{n=0}^{\infty} \delta^n S_n(x)}, \quad \delta \to 0
\end{equation}
where $\delta$ is a function of $\varepsilon$ to be determined later. Substituting this ansatz into Eq.~(\ref{stability03}) and dividing off the exponential factors, we get 
\begin{eqnarray}
\label{sing-term1}
&&\frac{\varepsilon}{\delta^4} [S_0'(x)]^4 + \frac{\varepsilon}{\delta^3} T_1(x) + \frac{\varepsilon}{\delta^2} T_2(x) + \frac{\psi}{\delta^2} x^{\alpha} [S_0'(x)]^2 \nonumber \\ &+& \frac{\psi}{\delta} T_3 (x) + \frac{\varepsilon}{\delta} T_4(x) + \dots = -1,
\end{eqnarray}
where only the singular terms in $\delta$ are written and where $T_i(x)$ are functions of $S_i(x)$ which are not explicitly given here for simplicity. By dominant balance in the limit $\delta \to 0$, the most singular term in $\delta$ must have the same order of magnitude than the constant term on the right hand side of Eq.~(\ref{sing-term1}) which imposes
\begin{equation}
\delta = \varepsilon^{1/4}.
\end{equation}
Substituting this last expression into Eq.~(\ref{sing-term1}) leads to 
\begin{eqnarray}
\label{sing-term2}
&&[S_0'(x)]^4 + \varepsilon^{1/4} T_1(x) + \varepsilon^{1/2} T_2(x) + \frac{\psi}{\varepsilon^{1/2}} x^{\alpha} [S_0'(x)]^2 \nonumber \\ &+& \frac{\psi}{\epsilon^{1/4}} T_3 (x) + \varepsilon^{3/4} T_4(x) + \dots = -1.
\end{eqnarray}
Again, by dominant balance in the limit $\varepsilon \to 0$, the most singular term in $\varepsilon$ must have the same order of magnitude than the constant term on the right hand side of Eq.~(\ref{sing-term2}) which imposes
\begin{equation}
\psi = [b(\alpha)]^{\alpha}\, \varepsilon^{1/2},
\end{equation}
where $b$ is a constant number of order 1. Returning to the definitions (\ref{eps-phi}) of $\varepsilon$ and $\psi$ we obtained the scalings we are searching for
\begin{equation}
\label{asymp-scaling}
\beta = b(\alpha)\, \eta^{\frac{1}{2\alpha}}.
\end{equation}
As shown in Fig.~\ref{fig01b}, these scalings agree with the numerical solution at large values of $\eta$ with $b(1)=2$ and $b(2)=3^{1/4}$. 

Using the definitions of $C_{\alpha}$, $\beta$ and $\eta$ (see Eqs.~(\ref{c-def}), (\ref{beta}) and (\ref{eta})), Eq.~(\ref{asymp-scaling}) gives the critical height of granular material, $L_{\text{c}}$, above which the system is unstable in terms of the control parameters of the system:
\begin{align}
\label{janssen-limit-scaling}
\text{Janssen limit:}& \quad L_{\text{c}}=\frac{2}{\sqrt{3(1-\nu^2)}}\frac{Et^2}{\rho_{\text{g}}g R^2}, \\
\label{hydro-limit-scaling}
\text{Hydrostatic limit:}& \quad L_{\text{c}}=\left[\frac{2 \xi}{\sqrt{(1-\nu^2)}}\frac{Et^2}{\rho_{\text{g}}g R}\right]^{1/2}.
\end{align}
Notice that we have neglected the constant terms found in the fits reported in Fig.~\ref{fig01b} since these terms are negligible when $\eta \gg 4$.

The experimental data reported in Figs. 3a, 3b and 3c of the main text show that $L_{\text{c}} \sim \rho^{-1/2}$, $L_{\text{c}} \sim t$ and $L_{\text{c}} \sim R^{-2/5}$. Consequently, the hydrostatic limit is the relevant regime in our experiment. The small discrepancy for the evolution of $L_{\text{c}}$ with $R$ is explained by the final expression~(\ref{critical-height-sup}) of $L_{\text{c}}$  derived in Sec.~\ref{sec:crit-height}. However, experimental data reported in Fig. 3d of the main text show also a dependence on the size $d$ of the granular material which is not captured by this analysis. This is the subject of the following sections.

\subsection{Imperfect shell}

\begin{figure}
\includegraphics[width=\columnwidth]{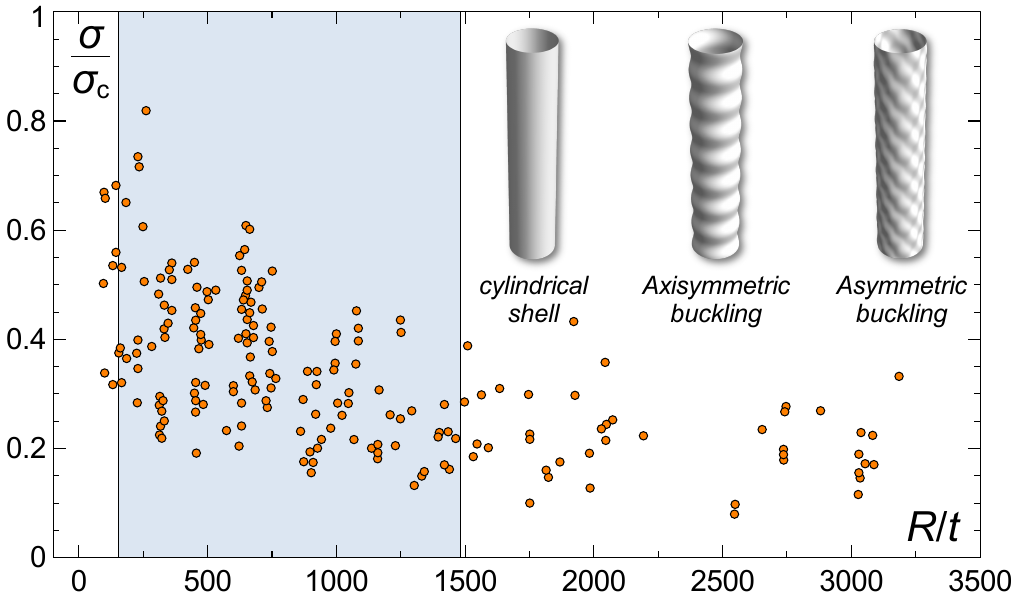}
\caption{Comparison between the experimental ($\sigma$) and theoretical ($\sigma_{\text{c}}$) critical stress for the buckling of axially compressed cylindrical shells as a function of the ratio $R/t$, where $R$ is the cylinder radius ($\nu=0.3$)~[12]. The shaded area shows the interval of values of the ratio $R/t$ considered in this work. Axisymmetric and asymmetric buckling modes are also shown.}
\label{fig02}
\end{figure}

As shown in Fig.~\ref{fig02}, early experimental tests indicated that real cylinders buckle at loads much lower than the classical buckling load (\ref{classical-stress})~[13, 14]. The discrepancy gets larger as the ratio $R/t$ increases. 

The search for reasons responsible for this discrepancy led to an enormous amount of research~[15]. This significant deviation from classical theory could, a priori, result from prebuckling deformations, geometric imperfections or load eccentricities. It was shown that the effect of prebuckling deformations is small and is not a primary reason for the difference between the classical prediction and experimental results and the great scatter of experimental results shown in Fig.~\ref{fig02}~[16, 17]. For axially compressed isotropic cylinders, small load eccentricities do not have either a major influence on the buckling strength~[18]. It was consequently shown that the single dominant factor contributing to the discrepancy between theory and experiment for axially compressed isotropic cylinders is geometric imperfections~[19].

An asymptotic formula, valid for small magnitude of the imperfections, has been developed in Ref.~[20] and tested in Ref.~[21]. This relation states that if there is an initial localized axisymmetric imperfection $w_0$ on the surface of a cylinder of constant thickness $t$, the stress, $\sigma$, for which the shell buckles under axial compression is given by:
\begin{equation}
\label{P-imperfect-sup}
\left(1-\frac{\sigma}{\sigma_{\text{c}}} \right)^{\frac{3}{2}} = \eta \frac{\sigma}{\sigma_{\text{c}}},
\end{equation}
where $\sigma_{\text{c}}$ is the classical buckling stress~(\ref{classical-stress}). The expression of $\eta$ is determined by the shape of the imperfection as follows
\begin{equation}
\label{Delta-sup}
\eta =\frac{3^{\frac{3}{2}}|\Delta|}{2} , \quad \Delta= \sqrt{\frac{1-\nu^2}{2}} \int_{-\infty}^{\infty} \frac{w_0(\tilde{x})}{h} e^{i \tilde{x}} d\tilde{x},
\end{equation}
with
\begin{equation}
\label{lambdac}
\tilde{x} = \frac{\pi x}{\lambda_{\text{c}}}, \quad \lambda_{\text{c}} = \frac{\pi \sqrt{Rt}}{[12(1-\nu^2)]^{\frac{1}{4}}},
\end{equation}
where $\lambda_{\text{c}}$ is the half wavelength of the classical axisymmetric buckling mode. Once the shape $w_0$ of the imperfection is given, $\Delta$ can be computed as well as $\sigma$. 

\begin{figure}
\includegraphics[width=\columnwidth]{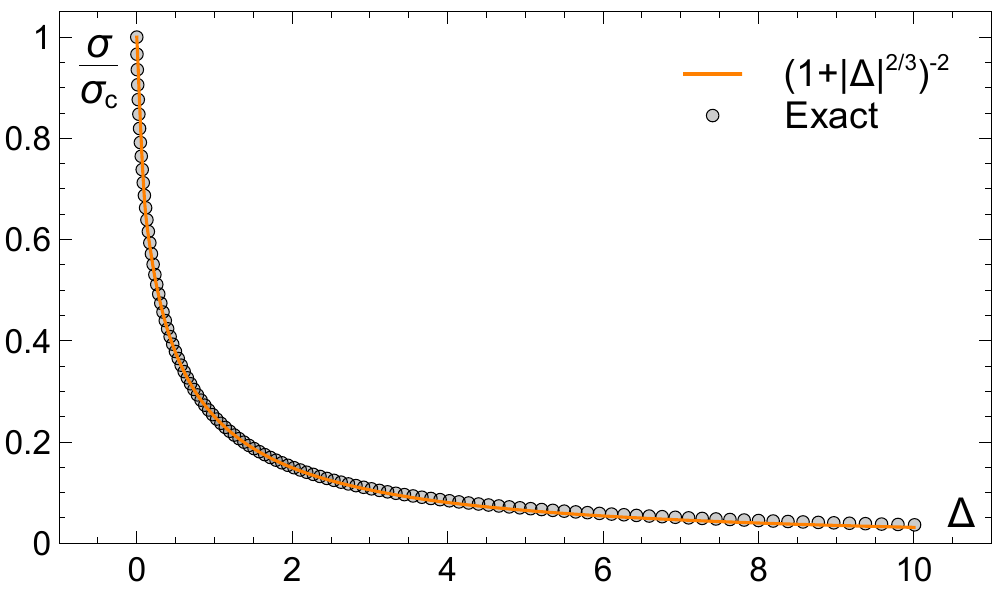}
\caption{Comparison between the exact solution of Eq.~(\ref{P-imperfect-sup}) and the approximation (\ref{sol-delta-sup}).}
\label{fig03}
\end{figure}

Equation (\ref{P-imperfect-sup}) is a third order algebraic equation in the variable $(\sigma/\sigma_{\text{c}})^{1/3}$ and can thus be solved exactly to obtain the evolution of $\sigma$ as a function of the imperfection parameter $\Delta$. However, the resulting cumbersome relation would not be so useful for the remaining analysis. Since Eq.~(\ref{P-imperfect-sup}) is an asymptotic formula valid when $\Delta$ is not too large, we use the following simple function which fits well the exact solution for $\Delta \in [0,10]$ as shown in Fig.~\ref{fig03}:
\begin{equation}
\label{sol-delta-sup}
\frac{\sigma}{\sigma_{\text{c}}}\simeq \left(1+|\Delta|^{\frac{2}{3}}\right)^{-2}.
\end{equation}

In order to compute $\Delta$, let's consider an imperfection, $w_0$, with a typical amplitude $\delta$ and which extend over a distance $\lambda_{\text{d}}$. In other words, we consider $w_0=\delta f(x)$, where $f(0)=1$, $|f(x)| \le 1$ and where $f(x)$ is vanishing for $|x| > \lambda_{\text{d}}$ or equivalently for $|\tilde{x}| > \pi \lambda_{\text{d}} / \lambda_{\text{c}}$. We expand Eq.~(\ref{Delta-sup}) in the first order in $\pi \lambda_{\text{d}} / \lambda_{\text{c}}$ and obtain
\begin{eqnarray}
\Delta &=& \sqrt{\frac{1-\nu^2}{2}} \frac{\delta}{t} \int_{-\pi \lambda_{\text{d}} / \lambda_{\text{c}}}^{\pi \lambda_{\text{d}} / \lambda_{\text{c}}} f(x) e^{i \tilde{x}} d\tilde{x} \\
\label{Delta-expan}
&\simeq & \sqrt{\frac{1-\nu^2}{2}} \frac{\delta}{t}\, \frac{2\pi \lambda_{\text{d}}}{\lambda_{\text{c}}} f(0) + O\left[\left(\frac{\pi \lambda_{\text{d}}}{\lambda_{\text{c}}}\right)^3\right] \\ 
\label{Delta-approx}
&=& \pi \sqrt{2(1-\nu^2)}\, \frac{\delta}{t} \frac{\lambda_{\text{d}}}{\lambda_{\text{c}}}.
\end{eqnarray} 
We thus obtain a universal expression for $\Delta$ which do not depend on the specific shape of the imperfection $w_0$ but only on its amplitude, $\delta$, and its spatial extension $\lambda_{\text{d}}$. 

\section{Influence of the grain size}

\subsection{Modified critical stress}

The ``fluid"-structure coupling is performed by assuming that the imperfections are induced by the granular material. We do the following assumptions:
\begin{equation}
\label{physical-assump}
\delta \sim t, \quad \text{and} \quad \lambda_{\text{d}} \sim d.
\end{equation}
The amplitude of the imperfection induced by the grains is of order $t$ and it spatial extension is of the order of the grain size. Using Eqs.~(\ref{Delta-approx}) and (\ref{physical-assump}) together with the expression (\ref{lambdac}) of $\lambda_{\text{c}}$ we get
\begin{equation}
\label{Delta-final-sup}
\Delta = \gamma \frac{d}{\sqrt{R t}}.
\end{equation}
where $\gamma$ is some numerical parameter fixed by the experimental data. Finally, using Eq.~(\ref{sol-delta-sup}) together with (\ref{Delta-final-sup}) we get the new expression of the critical stress above which the silo buckles
\begin{equation}
\label{sig-final}
\sigma =\frac{Et}{\sqrt{3(1-\nu^2)}R}\left[1+ \left(\frac{\gamma d}{\sqrt{R t}}\right)^{\frac{2}{3}}\right]^{-2}.
\end{equation}

\subsection{Critical height of granular material}
\label{sec:crit-height}

As already discussed in Sec.~\ref{sec:wkb}, the hydrostatic limit is the relevant regime in our experiments. The scaling (\ref{hydro-limit-scaling}) derived from an exact calculation of the silo stability in the limit of large $\eta$ (see Eq.~(\ref{eta})) can also be obtained, up to a constant multiplicative factor, from a simple and straightforward comparison between the classical critical stress $\sigma_c$ (\ref{classical-stress}) and the typical order of magnitude of granular load per unit section $F_{\mu}/(2\pi Rt)$ (see Eq.~(\ref{hydro-limit}) evaluated at $z=L$). Consequently, the scaling involving the finite size of granular material can be obtained in the same way
\begin{equation}
\sigma = \frac{2}{\sqrt{3}}\frac{F_{\mu}(L)}{(2\pi Rt)},
\end{equation}
where the numerical factor ensure that we recover the exact scaling (\ref{hydro-limit-scaling}) in the limit $d \to 0$ (perfect shell in our model). Using Eqs.~(\ref{hydro-limit}) and (\ref{sig-final}) we obtain
\begin{equation}
\label{critical-height-dim}
L_{\text{c}}=\left[\frac{2 \xi}{\sqrt{(1-\nu^2)}}\frac{Et^2}{\rho_{\text{g}} g R}\right]^{1/2} \left[1+ \left(\frac{\gamma d}{\sqrt{R t}}\right)^{\frac{2}{3}}\right]^{-1}
\end{equation}
Defining
\begin{equation}
\label{chi-def}
\bar{L}_{\text{c}} = \frac{L_{\text{c}}}{t} \sqrt{\frac{\rho g R}{E}}, \quad \bar{d}=\frac{d}{\sqrt{R t}}, \quad \chi=\frac{\sqrt{2\xi/\varphi}}{(1-\nu^2)^{1/4}},
\end{equation}
we finally obtain
\begin{equation}
\label{critical-height-sup}
\bar{L}_{\text{c}}=\frac{\chi}{1+ (\gamma \bar{d})^{\frac{2}{3}}}.
\end{equation}
This relation (\ref{critical-height-sup}) has been compared to rescaled experimental data in Fig.~3e of the main text with a good agreement provided
\begin{equation}
\label{chi-gamma}
\chi = 5.0 \pm 0.2 \quad \text{and} \quad \gamma = 0.11 \pm 0.03.
\end{equation}
Since $\bar{d} <6.5$ in our experiments (see Fig.~3e of the main text), this small value of $\gamma$ ensures that $\Delta$, defined in Eq.~(\ref{Delta-final-sup}), is always smaller than 1 and justifies the use of an asymptotic theory for imperfect shell as well as the validity of the approximation (\ref{sol-delta-sup}) and the expansion (\ref{Delta-expan}). 

\subsection{Discussion about the hydrostatic limit}
\label{sec:disc-hydro}

\begin{figure}
\includegraphics[width=\columnwidth]{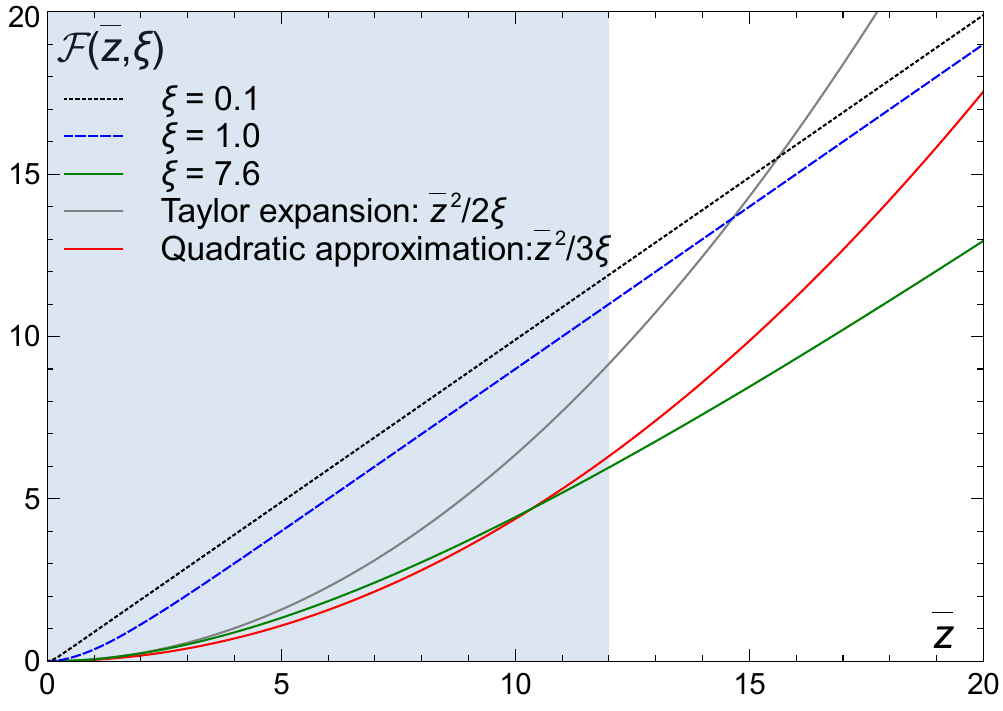}
\caption{Plots of the function ${\cal F}(\bar{z},\xi)$ for several value of $\xi$ and $\bar{z}$ varying over the relevant range characterizing our experiments. The Taylor expansion (\ref{hydro-limit}) together with a quadratic approximation of the function ${\cal F}(\bar{z},\xi)$ are also plotted for $\xi=7.6$ which is the typical value in our experiments.}
\label{fig04}
\end{figure}

The difference between the dependence of $L_{\text{c}}$ on the control parameters reported in Eqs.~(\ref{janssen-limit-scaling}) and (\ref{hydro-limit-scaling}) is due to the linear and quadratic dependence of $F_{\mu}$ with respect to the coordinate $z$ in the two asymptotic regimes defined by actual value of $\xi$, see Eqs.~(\ref{janssen-limit}) and (\ref{hydro-limit}). These two asymptotic regimes are obtained by expanding the function ${\cal F}(\bar{z},\xi)$, defined in Eq.~(\ref{f-def}), for $\xi \ll 1$ or $\xi \gg L/R$. As long as the function ${\cal F}(\bar{z},\xi)$ can be approximated by a quadratic function of $z$, the relevant scaling for $L_{\text{c}}$ is the one obtained in Eq.~(\ref{hydro-limit-scaling}) and which describes well our experiments. 

\begin{figure}
\includegraphics[width=\columnwidth]{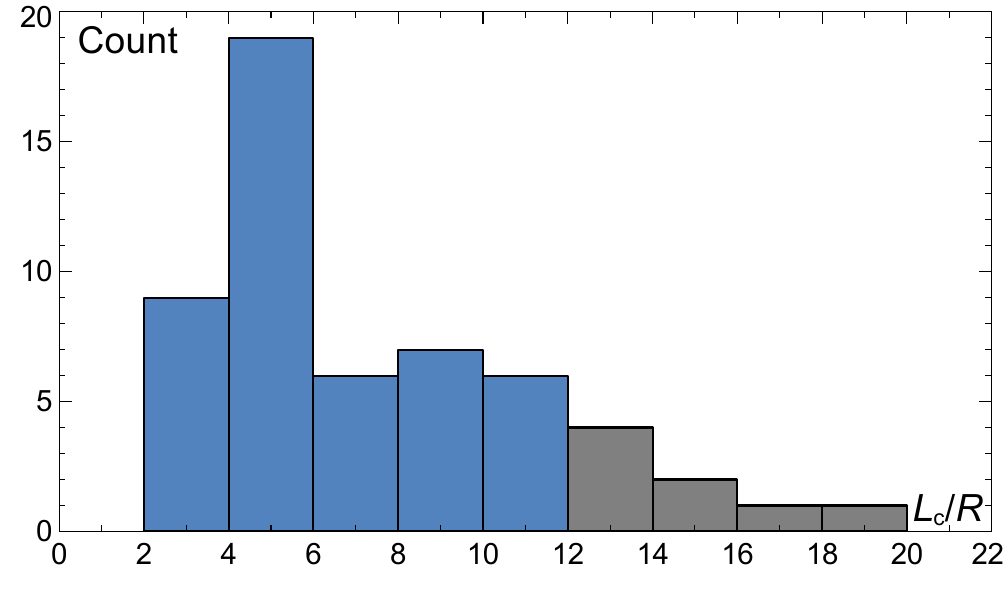}
\caption{Distribution of the ratio $L_{\text{c}}/R$ in the experiments.}
\label{fig05}
\end{figure}

From the range of values of $\chi$ obtained by fitting the experimental data (\ref{chi-gamma}) and the relation (\ref{chi-def}) between $\chi$ and $\xi$, we obtain
\begin{equation}
\label{xi-val}
\xi = 7.6 \pm 0.8,
\end{equation}
for a packing fraction $\varphi \simeq 0.64$. As shown in Fig.~\ref{fig04}, for $\xi \lesssim 1$, the function ${\cal F}(\bar{z},\xi)$ is essentially linear in $\bar{z}=z/R$ except for very small values of $\bar{z}$. However, for a value of $\xi$ compatible with our experiments, we note that the function ${\cal F}(\bar{z},\xi)$ can be well approximated by a quadratic function provided the coefficient of the Taylor expansion (\ref{hydro-limit}) is slightly adjusted which is equivalent to a redefinition of $\xi$. This means that even if $\xi$ is not much larger than $L_{\text{c}}/R$ to fully justify the Taylor expansion (\ref{hydro-limit}) characterizing the hydrostatic limit, $\xi$ is large enough in our experiments to allow the system to be described by an effective hydrostatic regime where ${\cal F}(\bar{z},\xi)$ can be well approximated by a quadratic function of $\bar{z}$. However, as shown in Fig.~\ref{fig04}, this effective hydrostatic regime breaks down for $\bar{z} \gtrsim 12$. Since $0\le \bar{z} \le L_{\text{c}}/R$, it implies that the hydrostatic approximation is not suitable for experimental data for which $L_{\text{c}}/R \gtrsim 12$. Those data are marked by a cross in Fig. 3e of the main text. Figure~\ref{fig05} shows the distribution of $L_{\text{c}}/R$ obtained in the experiments showing that most of them lie on the region $L_{\text{c}}/R \le 12$ which justifies the use of an effective hydrostatic regime to describe the data.

A simple criteria can be applied to determine when this effective hydrostatic regime takes place. For a given value of $\xi$, one can fit the function ${\cal F}(\bar{z},\xi)$ by a quadratic function $a \bar{z}^2$ for $\bar{z} \in [0,L/R]$ and compute the coefficient of determination $R^2$ as a function of $L/R$. $R^2$ is of course a decreasing function of $L/R$. One can then search the value of $L/R=(L/R)_{\text{max}}$ such $R^2 = R_0^2$, where $R_0^2$ is a value of the coefficient of determination close enough to 1 to ensure a good approximation. Consequently, for that particular value of $\xi$, a quadratic function is a good approximation for $L/R \le (L/R)_{\text{max}}$. By varying $\xi$, one can determine $(L/R)_{\text{max}}$ as a function of $\xi$. It appears that this relation is linear:
\begin{equation}
\label{const-lr}
\left(\frac{L}{R}\right)_{\text{max}} = \Omega(R_0^2) \, \xi
\end{equation}
where $\Omega(0.995)\simeq 9/5$, $\Omega(0.996)\simeq \pi/2$ and $\Omega(0.997)\simeq 4/3$. Consequently, while a hydrostatic regime takes place only for $L/R \ll \xi$, an effective hydrostatic regime applies for $L/R$ as large as $\sim 1.6\, \xi$. 

\begin{figure}
\includegraphics[width=\columnwidth]{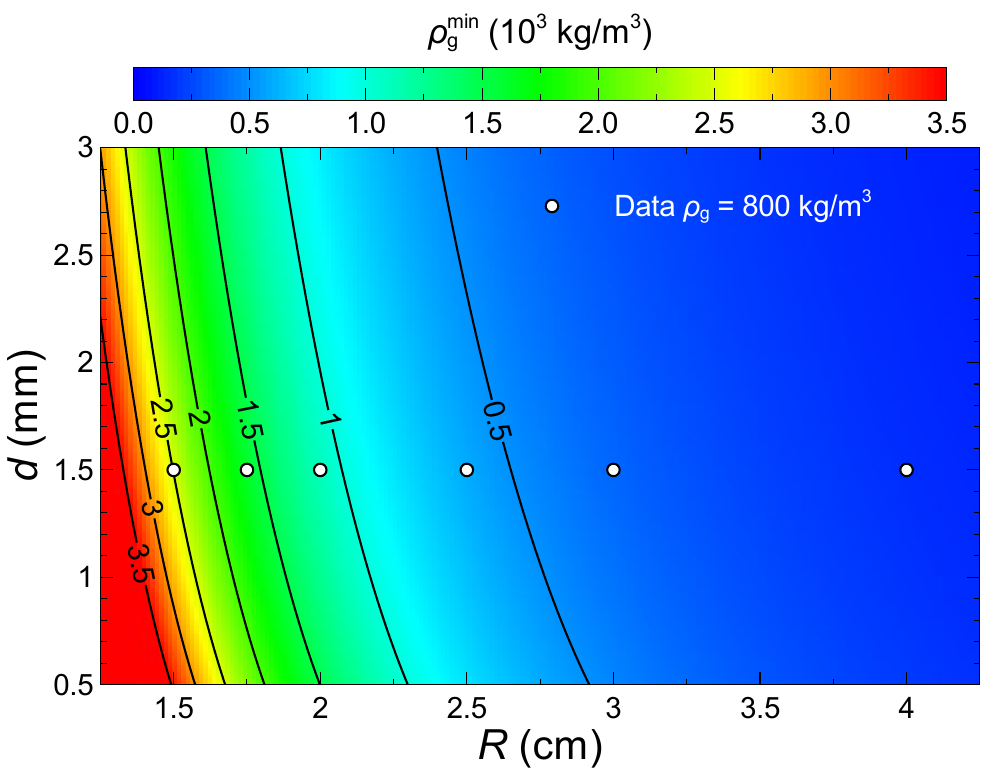}
\caption{Evolution of $\rho_{\text{g}}^{\text{min}}$ defined in Eq.~(\ref{const-rho}) as a function of $d$ and $R$ for $t=27 \mu$m, $E=2$ GPa, $\Omega = \pi/2$, $\xi=7.6$ and $\gamma = 0.11$. Data with $\rho_{\text{g}}=800$ kg/m$^3$ reported in Fig. 3c of the main text are also displayed.}
\label{fig06}
\end{figure}

The constrain (\ref{const-lr}) together with the expression (\ref{critical-height-sup}) of $L_{\text{c}}$ gives the sub-domain of the parameter space where the effective hydrostatic regime applies. For example, this relation among the control parameters of the system can be written as a constrain on the density of the granular medium as follows
\begin{equation}
\label{const-rho}
\rho_{\text{g}} > \frac{2}{\Omega^2 \, \xi} \frac{Et^2}{gR^3} \left[1+\left(\gamma \frac{d}{\sqrt{R t}}\right)^{\frac{2}{3}}\right]^{-2} \equiv \rho_{\text{g}}^{\text{min}}
\end{equation}
As an example, Fig.~\ref{fig06} shows the evolution of $\rho_{\text{g}}^{\text{min}}$ as a function of $d$ and $R$. The data with $\rho_{\text{g}}=800$ kg/m$^3$ reported in Fig. 3c of the main text are also displayed showing that, for $R\le 2$ cm, the density is too low to allow the effective hydrostatic regime to take place.

The constrain (\ref{const-lr}) can be recovered by another route revealing its physical meaning. The shear force applied to the wall in the pseudo hydrostatic regime should be smaller than the shear force in the Janssen limit. Consequently, the collapse length (\ref{hydro-limit-scaling}) computed in the hydrostatic regime, noted here $L_{\text{c}}^{(\text{H})}$, should be larger than the collapse length (\ref{janssen-limit-scaling}) computed in the Janssen regime and noted here $L_{\text{c}}^{(\text{J})}$: $L_{\text{c}}^{(\text{H})}>L_{\text{c}}^{(\text{J})}$. From Eqs.~(\ref{janssen-limit-scaling}) and (\ref{hydro-limit-scaling}) we also have $L_{\text{c}}^{(\text{J})} = (L_{\text{c}}^{(\text{H})})^2 / \sqrt{3} \xi R$. Combining these two relations we obtain $L_{\text{c}}^{(\text{H})} < \sqrt{3} \xi R$ which is equivalent to Eq.~(\ref{const-lr}) with almost the same numerical factor. Thus, the pseudo hydrostatic regime applies when the shear force can be approximated by a quadratic function with sufficiently good accuracy or equivalently when the corresponding collapse height is smaller then the collapse height obtained in the Janssen limit.

\begin{figure}[b]
\includegraphics[width=\columnwidth]{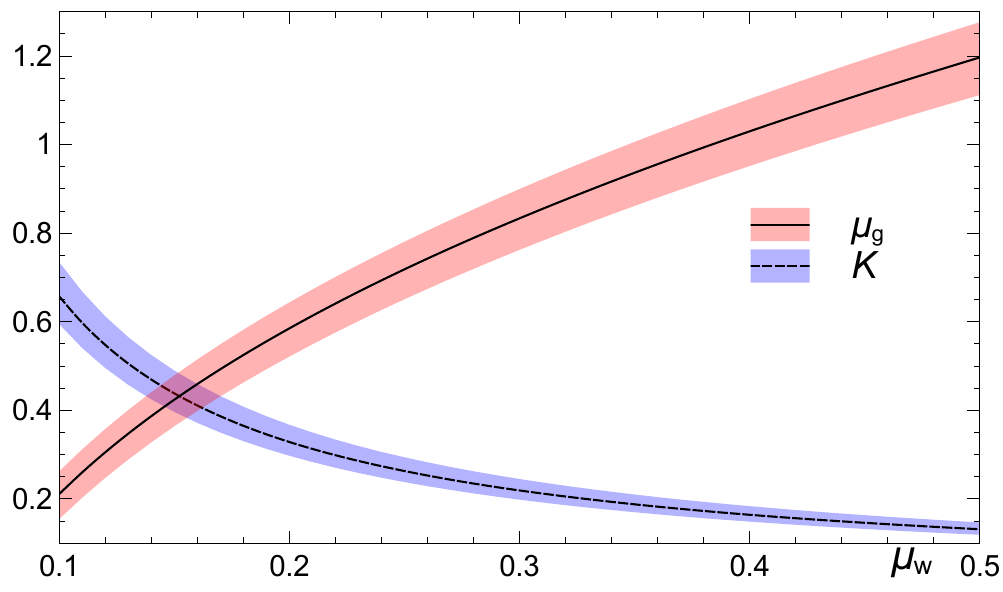}
\caption{Evolution of $\mu_{\text{g}}$ and $K$, defined in Eq.~(\ref{mug}), as a function of the grain-wall friction coefficient $\mu_{\text{w}}$. The shaded areas show the region spanned when $\xi$ varies in the interval (\ref{xi-val}).}
\label{fig07}
\end{figure}

Finally, the values of $\xi$ characterizing our experiments (\ref{xi-val}) can be related to the friction coefficient between the wall and the grains, $\mu_{\text{w}}$, and the friction coefficient between the grains, $\mu_{\text{g}}$. If $\sigma_{rr}$ (see Sec.~\ref{sec:gran-load}) is the minor principal stress (active case), we have~[22, p. 84]
\begin{equation}
\label{def-k}
K=\frac{1-\sin \phi}{1+\sin \phi}, \quad \mu_{\text{g}} = \tan \phi,
\end{equation}
$\phi$ being the angle of friction. Using the definition (\ref{sigmazz}) of $\xi$ and Eq.~(\ref{def-k}), we find
\begin{equation}
\label{mug}
\mu_{\text{g}}=\frac{2\mu_{\text{w}}\xi-1}{2\sqrt{2\mu_{\text{w}}\xi}}, \quad K=\frac{1}{2\mu_{\text{w}}\xi}.
\end{equation}
The evolution of $\mu_{\text{g}}$ and $K$ as a function of the grain-wall friction coefficient $\mu_{\text{w}}$ is given in Fig.~\ref{fig07} using the values of $\xi$ characterizing our experiments.

\subsection{Critical thickness $t_{\text{c}}$}

One striking feature of the experimental data reported in Fig. 3b of the main text, where the evolution of $L_{\text{c}}$ is plotted as a function of the silo thickness $t$, is that a linear extrapolation of the data toward $L_{\text{c}} \to 0$ leads to an apparent finite critical thickness around 10 $\mu$m. In the limit $L_{\text{c}} \to 0$, the granular load vanishes and the remaining load applying on the silo wall is its own weight. However, the typical height of the paper silo used in the experiment is 40 cm. Such a silo with a thickness of 10 $\mu$m should not collapse under its own weight. Indeed, the comparison of the critical stress (\ref{sig-final}) with the average self-weight force (\ref{self-weight}) per unit section leads to
\begin{equation}
\frac{Et}{\sqrt{3(1-\nu^2)}R}\left[1+ \left(\frac{\gamma d}{\sqrt{R t}}\right)^{\frac{2}{3}}\right]^{-2} = \rho_{\text{w}} g H.
\end{equation}
In the limit of vanishing thickness, one finds
\begin{equation}
t_{\text{c}}=[3(1-\nu^2)]^{3/10} \left(\frac{\rho_{\text{w}} g H}{E}\right)^{3/5} (\gamma d)^{4/5} R^{1/5}.
\end{equation}
With the parameters used in the experiments reported in Fig. 3b of the main text ($d=3$ and $4.5$ mm, $R=2$ and $2.55$ cm, $H=40$ cm, $\gamma=0.11$, see also Table~\ref{table1}), one finds $0.30$ $\mu$m $<t_{\text{c}} < 0.41$ $\mu$m. This is one order of magnitude smaller than the thickness found by linear extrapolation (of course, using the critical stress (\ref{classical-stress}) instead of (\ref{sig-final}) leads to even smaller critical thickness). 

Actually, the theoretical curve describing the data (see dashed lines in Fig. 3b of the main text) is convex for small $L_{\text{c}}$ leading thus to a smaller value of $t_{\text{c}}$ than the one obtained from a linear extrapolation. The apparent critical thickness can actually be obtained directly from Eq.~(\ref{critical-height-dim}). For this purpose, we apply the same procedure to the theoretical expression than the one used for the experimental data. We consider the asymptotic behavior of Eq.~(\ref{critical-height-dim}) and search for which value of $t$ it vanishes (which is equivalent to extrapolate to vanishing $L_{\text{c}}$). With $q= (\gamma d)^2/R$, Eq.~(\ref{critical-height-dim}) can be written as follow together with its asymptotic expansion
\begin{equation}
\label{critical-asymp-height-sup}
L_{\text{c}}= \frac{p t}{\left[1+ \left(\frac{q}{t}\right)^{\frac{1}{3}}\right]} \underset{t\gg q}{\simeq} p \left(t^{\frac{2}{3}}+q^{\frac{2}{3}}\right)\left(t^{\frac{1}{3}}-q^{\frac{1}{3}}\right) 
\end{equation}
The asymptotic expansion vanishes at a value of $t = q$ corresponding to the apparent critical thickness obtained by extrapolating linearly the experimental data:
\begin{equation}
t_{\text{c}}^{\text{apparent}} = \frac{(\gamma d)^2}{R}.
\end{equation}
With the parameters used in the experiments reported in Fig. 3b of the main text, one finds 
\begin{equation}
1 \, \mu\text{m} <t_{\text{c}}^{\text{apparent}} < 12\, \mu\text{m}
\end{equation}
in good agreement with the value found by extrapolating the experimental data.

\vspace{1cm}
\hrule

\begin{enumerate}[label={[\arabic*]},noitemsep]
\item H. A. Janssen, Zeitschrift des Vereins Deutscher Ingenieure {\bf 39}, 1045 (1895); English translation: M. Sperl, Granular Matter {\bf 8}, 59 (2006).
\item G. D. Scott, Nature {\bf 188}, 908 (1960).
\item G. D. Scott, Nature {\bf 194}, 956 (1962).
\item J. D. Bernal, Proc. R. Soc. London Ser. A {\bf 280}, 299 (1964).
\item J. L. Finney, Proc. R. Soc. London Ser. A {\bf 319}, 479 (1970).
\item R. Lorenz, Z. Ver. Deut. Ingr. {\bf 52}, 1766 (1908).
\item S. P. Timoshenko, Z. Math. Physik {\bf 58}, 337 (1910).
\item R. Lorenz, Physik. Z. {\bf 13}, 241 (1911).
\item R. V. Southwell, Phil. Trans. Roy. Soc. London, Series A {\bf 213}, 187 (1914).
\item S. P. Timoshenko and J. M. Gere, {\it Theory of elastic stability}, 2nd edition, McGraw-Hill, 1961.
\item C. M. Bender and S. A. Orszag, {\it Advanced Mathematical Methods for Scientists and Engineers}, McGraw-Hill, 1978.
\item D. Bushnell, AIAA Journal {\bf 19}, 1183 (1981).
\item W. Flugge, Ingenieur-Archiv {\bf 3}, 463 (1932).
\item E. E. Lundquist, NACA Tech Note, No 473 (1933).
\item J. G. Teng, Appl. Mech. Rev. {\bf 49}, 263 (1996).
\item D. J. Gorman and R. M. Evan-Iwanowski, Dev. Theor. Appl. Mech. {\bf 4}, 415 (1970).
\item N. Yamaki and S. Kodama, Report of the Inst of High Speed Mech {\bf 25}, Tohoku Univ, 99 (1972).
\item G. J. Simitses, D. Shaw, I. Sheinman and J. Giri, Composites Sci. Tech. {\bf 22}, 259 (1985).
\item G. J. Simitses, Appl. Mech. Rev. {\bf 39}, 1517 (1986).
\item J. C. Amazigo and B. Budiansky, J. Appl. Mech. {\bf 39}, 179 (1972).
\item J. W. Hutchinson, R. C. Tennyson and D. B. Muggeridge, AIAA Journal {\bf 9}, 48 (1971).
\item R. M. Nedderman, {\it Statics and Kinematics of Granular Materials}, Cambridge University Press, 1992.
\end{enumerate}

\end{document}